\documentclass[aps,prl,twocolumn,showpacs]{revtex4}

\newcommand{\comp}{
            {\rm C}\llap{\vrule height7.1pt width1pt depth-.4pt\phantom t}}

\begin{document}

\preprint{UFIFT-HEP-02-04,CRETE-02-11}

\title{Plane waves in a general Robertson-Walker background}

\author{N. C. Tsamis}
\email[]{tsamis@physics.uoc.gr}
\affiliation{Department of Physics, University of Crete, 
             GR-71003 Heraklion, Greece}

\author{R. P. Woodard}
\email[]{woodard@phys.ufl.edu}
\affiliation{Department of Physics, University of Florida,
             Gainesville, FL 32611, USA}

\date{\today}

\begin{abstract}

We present an exact solution for the plane wave mode functions of a
massless, minimally coupled scalar propagating in an arbitrary homogeneous,
isotropic and spatially flat geometry. Our solution encompasses all previous
solvable special cases such as de Sitter and power law expansion. Moreover,
it can generate the mode functions for gravitons. We discuss some of the 
many applications that are now possible. 

\end{abstract}


\maketitle

{\it 1. Introduction:} The universe is homogeneous and isotropic on the 
largest observable scales \cite{KT}. Recent precision measurements of 
anisotropies in the cosmic microwave background strongly support the idea 
that it is also spatially flat \cite{MAXIMA,BOOMERANG,WMAP}. This is in any 
case the theoretical prejudice of inflationary cosmology \cite{Linde}. 
The invariant element of such a universe can be simply expressed in 
co-moving coordinates,
\begin{equation}
ds^2 = -dt^2 + a^2(t) d\vec{x} \cdot d\vec{x} \; . \label{HI}
\end{equation}
Although the scale factor $a(t)$ is not measurable, its derivatives
can be formed to give observables,
\begin{equation}
H(t) \equiv \dot{a} \, a^{-1} \qquad , \qquad q(t) \equiv -1 - \dot{H} H^{-2} 
\; . \label{Hq}
\end{equation}
The Hubble parameter $H(t)$ is the cosmological expansion rate
while the deceleration parameter $q(t)$ provides a measure of the rate
at which this expansion is slowing down ($q > -1$) or speeding up
($q < -1$).

It has long been realized that an expanding universe can result in 
particle creation \cite{Parker1}. A particularly interesting and 
likely candidate for this effect is the massless, minimally coupled
scalar,
\begin{equation}
{\cal L} = -\frac12 \partial_{\mu} \varphi \partial_{\nu} \varphi g^{\mu\nu} 
\sqrt{-g} \; .
\end{equation}
Because these particles are massless without classical conformal invariance,
there can be an appreciable amplitude for virtual pairs to appear with
wave lengths comparable to $H^{-1}$. When this happens the pairs are 
pulled apart by the expansion of spacetime and there is physical 
particle creation \cite{Parker2}. When a scalar potential is added to 
drive inflation, essentially this mechanism \cite{Slava} is the leading 
candidate for explaining the primordial cosmological perturbations which 
seeded structure formation and whose imprint is observed in the anisotropies 
of the cosmic microwave background \cite{infconf}.

An even more direct application derives from the observation 
\cite{Grishchuk1} that linearized gravitons obey the same equation of motion 
in the background (\ref{HI}). This leads to a fundamental tensor component 
in the primordial spectrum of cosmological perturbations \cite{MFB,LL,Rocky}.
There is also the possibility that quantum gravitational back-reaction
slows inflation \cite{TW1}. Although standard inflationary cosmology is 
realized via a scalar field, it is probably not natural to have interacting 
scalars which are much lighter than the Hubble parameter during inflation. 
In that respect the graviton can achieve all the advantages of inflation 
in a far more economical and natural way \cite{TW1}.

It is therefore frustrating that the scalar mode functions are only known 
for certain choices of $a(t)$ \cite{BD}. The fact that typical inflationary 
backgrounds are not of this type means that either perturbative or numerical
techniques must be used to predict the spectrum of cosmological perturbations 
\cite{MFB,LL,Rocky}. The absence of general mode solutions is most noticeable 
when back-reaction becomes important. By working in de Sitter background one 
can see that the slowing effect eventually becomes strong \cite{TW1}, but it 
is not possible to reliably follow where the physics is trying to lead.

The purpose of this paper is to solve the scalar mode equation {\it for any} 
scale factor $a(t)$. Section 2 gives the equation and defines what we mean 
by the ``mode function''. In Section 3 we present a solution which is valid 
for any period during which the deceleration parameter (\ref{Hq}) is constant. 
In Section 4 these solutions are extended to cover general evolution by 
representing it as a sequence of ever finer steps between regions of constant 
$q(t)$. The transfer matrix which makes this possible is expressed as the
path-ordered product of the exponential of a line integral. In Section 5 we
improve the constant deceleration solution by explicitly extracting the 
determinant of this matrix for general $a(t)$. We also identify and normalize 
the positive frequency solution. Expansions for the ultraviolet and infrared 
regimes are derived in Sections 6 and 7, respectively. Our results are 
collected in Section 8 and expressed in terms of standard parameters. We also 
discuss applications, including an improved result for the gravitational 
wave contribution to the power spectrum of anisotropies in the cosmic 
microwave background. 

{\it 2. The scalar mode equation:} We take advantage of spatial translation 
invariance to Fourier transform in the $(D \!-\!1)$-dimensional spatial 
coordinates,
\begin{equation}
\widetilde{\varphi}(t,\vec{k}) \equiv \int d^{\scriptstyle D \! - \!1}{\! x} 
\; e^{-i \vec{k} \cdot \vec{x}} \varphi(t,\vec{x}) \; .
\end{equation}
The equation of motion of $\widetilde{\varphi}(t,\vec{k})$ is,
\begin{equation}
\ddot{\widetilde{\varphi}} + (D \!-\! 1) H \dot{\widetilde{\varphi}} + k^2
a^{-2} \widetilde{\varphi} = 0 \; . \label{phieqn}
\end{equation}
The middle term, $(D \!-\! 1) H \dot{\widetilde{\varphi}}$, is responsible for 
the phenomenon of ``Hubble friction'' in which the expansion of spacetime
retards the scalar's evolution.

Our work was motivated by a recent paper of Finelli, Marozzi, Vacca and
Venturi \cite{FMVV} who studied the case of a minimally coupled scalar with 
arbitrary mass. We follow them in the standard step of extracting a factor of 
$a^{-\!\frac{D \!-\!  1}2}$ to eliminate the Hubble friction,
\begin{equation}
\widetilde{\varphi}(t,\vec{k}) = a^{-\!\frac{D \!-\! 1}2} \Bigl\{ u(t,k) \,
\alpha(\vec{k}) + u^*(t,k) \, \alpha^{\dagger}(-\vec{k}) \Bigr\} \; . 
\label{phitilde}
\end{equation}
The quantities $\alpha(\vec{k})$ and $\alpha^{\dagger}(-\vec{k})$ are time
independent quantum operators. What remains is the $\comp$-number mode 
function $u(t,k)$ and its conjugate. It obeys,
\begin{equation}
\ddot{u}(t,k) + H^2 \left[({\scriptstyle {k \over a H}})^2 
- ({\scriptstyle \frac{D\!-\!1}2}) {\scriptstyle \frac{\dot{H}}{H^2}} - 
({\scriptstyle \frac{D\!-\!1}2})^2 \right] u(t,k) = 0 \; . \label{diffeqn}
\end{equation}

To fix the normalization, suppose the creation and annihilation operators
commute canonically,
\begin{equation}
\left[ \alpha(\vec{k}) , \alpha^{\dagger}(\vec{k}') \right] = 
(2\pi)^{D \!-\! 1} \delta^{D \!-\! 1}(\vec{k} - \vec{k}') \; .
\end{equation}
Taking the time derivative of (\ref{phitilde}) gives,
\begin{eqnarray}
\lefteqn{\left[ \widetilde{\varphi}(t,\vec{k}) , \dot{\widetilde{\varphi}}(t,
\vec{k}') \right] = \left({ 2\pi \over a(t) }\right)^{D \!-\! 1} 
\delta^{D \!-\! 1}(\vec{k} + \vec{k}') } \nonumber \\
& & \hspace{1.5cm} \times \Bigl\{ u(t,k) \, \dot{u}^*(t,k) - u^*(t,k) \,
\dot{u}(t,k) \Bigr\} \; . \qquad \label{1stcom}
\end{eqnarray}
Since the momentum canonically conjugate to $\varphi(t,\vec{x})$ is $a^{D \!-\! 
1}(t) \dot{\varphi}(t,\vec{x})$, we must also have,
\begin{equation}
\left[ \widetilde{\varphi}(t,\vec{k}) , \dot{\widetilde{\varphi}}(t,\vec{k}')
\right] = i \left({ 2\pi \over a(t) }\right)^{D \!-\! 1} \delta^{D \!-\!
1}(\vec{k} + \vec{k}') \; . \label{2ndcom}
\end{equation}
Comparing (\ref{1stcom}) with (\ref{2ndcom}) determines the Wronskian of the
mode function,
\begin{equation}
u(t,k) \, \dot{u}^*(t,k) - u^*(t,k) \, \dot{u}(t,k) = i \; . \label{norm}
\end{equation}

{\it 3. The solution for constant deceleration:} Just as in \cite{FMVV}, we 
organize the dependence upon time and wave number into two dimensionless 
parameters,
\begin{equation}
x(t,k) \equiv {k \over a H} \qquad , \qquad y(t) \equiv -{\dot{H} \over H^2} 
\; .
\end{equation}
The meaning of $x(t,k)$ is the physical wave number in Hubble units. 
Super-horizon modes have $x < 1$ whereas sub-horizon modes have $x > 1$. 
The quantity $y(t)$ is related to the deceleration parameter, $q(t) = -1 + 
y(t)$. The Weak Energy Condition implies $y(t) \geq 0$. Slow roll inflation 
is characterized by $y(t) \ll 1$. 

Note that $y(t)$ is constant whenever $a(t)$ is a power of any linear function
of $t$. For example, a radiation dominated universe has $y = 2$; the result 
for a matter dominated universe is $y = \frac32$; and curvature domination
gives $y = 1$. Of course (\ref{diffeqn}) is trivially reducible to Bessel's 
equation whenever the scale factor obeys a power law, but the {\it order} of
the necessary Bessel functions varies with the power. This is why it has been
so difficult to find a general solution. In \cite{FMVV}, however, it was
recognized that {\it the general solution should involve a Bessel function 
whose order depends upon} $y(t)$. They worked only to first order in $y$ 
but it is straightforward to extend their result to all orders. The required 
generalization is,
\begin{equation}
{\cal H}(x,y) \equiv x^{\mu} H^{(1)}_{\nu}(\lambda x) \; , \label{consol}
\end{equation}
where the parameters are the following functions of $y$,
\begin{equation}
\mu \equiv {-\frac12 y \over 1 - y} \;\; , \;\; \nu \equiv {\frac{D\!-\!1}2 -
\frac12 y \over 1 - y} \;\; , \;\; \lambda \equiv {1 \over 1 - y} \; .
\end{equation}
Although the parameter $x$ is always positive, $\lambda$ passes from $+\infty$
to $-\infty$ as $y$ evolves from $1^-$ to $1^+$, which signals the end of
inflation. This does not lead to wild oscillations in ${\cal H}(x,y)$ because 
$\nu = \frac12 + (\frac{D}2  -1) \lambda$ makes the same transition, and the 
factor of $1/\Gamma(n + \nu + 1)$ in the series expansion of the Hankel 
function compensates the growth of $\lambda$. However, the possibility of 
$\lambda$ becoming negative does raise the issue of how we define $(\lambda 
x)^{\nu}$. Throughout this paper we shall understand negative $\lambda$ to 
mean $e^{+i \pi} \vert \lambda \vert$, so that $(- \vert \lambda \vert 
x)^{\nu} \equiv e^{i \nu \pi} (\vert \lambda \vert x)^{\nu}$.

To see that (\ref{consol}) solves (\ref{diffeqn}) for constant $y$, first
compute the (general) time derivatives of $x(t,k)$ and $y(t)$,
\begin{eqnarray}
\dot{y} & = & H \left[ 2 {\dot{H}^2 \over H^4} - {\ddot{H} \over H^3} 
\right] \; , \\
\dot{x} & = & - {k \over a} \left[ 1 + {\dot{H} \over H^2} \right] 
= - H x (1-y) \; , \label{xdot} \\
\ddot{x} & = & {H k \over a} \left[1 + {\dot{H} \over H^2} + 2 {\dot{H}^2 
\over H^4} - {\ddot{H} \over H^3} \right] \nonumber \\
& & \hspace{1cm} = H^2 x \Bigl(1 - y + {\dot{y} \over H} \Bigr) \; . 
\label{xddot}
\end{eqnarray}
Now use Bessel's equation to derive an identity for the second $x$ derivative 
of ${\cal H}(x,y)$,
\begin{eqnarray}
{\cal H}_{,xx} & = & \left({2 \mu - 1 \over x}\right) {\cal H}_{,x} -\lambda^2 
{\cal H} + \left({\nu^2 - \mu^2 \over x^2}\right) {\cal H} \; , \qquad \\
& = & - {{\cal H}_{,x} \over x (1-y)} - \left( {x^2 + \frac{D\!-\!1}2 y - 
(\frac{D\!-\!1}2)^2 \over x^2 (1 - y)^2} \right) {\cal H} \; . \label{H,xx}
\end{eqnarray}
When $\dot{y} = 0$ we have,
\begin{eqnarray}
\dot{\cal H} \Bigl\vert_{\dot{y}=0} & = & \dot{x} \, {\cal H}_{,x} \; , \\
\ddot{\cal H} \Bigl\vert_{\dot{y}=0} & = & \ddot{x} \, {\cal H}_{,x} + 
\dot{x}^2 \, {\cal H}_{,xx} \; , \\
& = & H^2 x (1-y) \, {\cal H}_{,x} + H^2 x^2 (1-y)^2 \, {\cal H}_{,xx} 
\; , \qquad \\
& = & - H^2 \left[x^2 + {\scriptstyle \frac{D\!-\!1}2} y - 
({\scriptstyle \frac{D\!-\!1}2})^2\right] {\cal H} \; .
\end{eqnarray}

{\it 4. The transfer matrix:} Let us follow Grishchuk \cite{Grishchuk2} in 
joining solutions between two periods of constant $y$. Suppose there is a
sudden transition at $t=t_1$, that $y(t) = y_1$ for $t < t_1$, and that $y(t) 
= y_2$ for $t > t_1$. The solution during each period must be a linear 
combination of the appropriate constant $y$ solutions,
\begin{eqnarray}
u_1(t,k) & = & c_1 {\cal H}(x,y_1) + d_1 {\cal H}^*(x,y_1) \quad \forall \; 
t < t_1 , \qquad \\
u_2(t,k) & = & c_2 {\cal H}(x,y_2) + d_2 {\cal H}^*(x,y_2) \quad \forall \; 
t > t_1 . \qquad 
\end{eqnarray}
Suppose further that the scale factor is continuous at the transition. Since 
the differential equation has no factors of $\delta'(t - t_1)$ its solution
must also be continuous at $t_1$,
\begin{equation}
c_2 {\cal H}_{12} + d_2 {\cal H}^*_{12} = c_1 {\cal H}_{11} + 
d_1 {\cal H}^*_{11} \; ,
\end{equation}
where we have compressed the notation by making the definition, ${\cal H}_{ij} 
\equiv {\cal H}\Bigl(x(t_i),y_j\Bigr)$. Suppose finally that the Hubble
constant is continuous. Since the differential equation has no factors of
$\delta(t - t_1)$ the first derivative of its solution must be continuous at 
the transition,
\begin{eqnarray}
\lefteqn{(1 - y_2) \Bigl[c_2 {\cal H}_{12 , x} + d_2 {\cal H}^*_{12 , x} 
\Bigr]} \nonumber \\
& & \hspace{2cm} = (1 - y_1) \Bigl[ c_1 {\cal H}_{11} + d_1 {\cal H}^*_{11 , x} 
\Bigr] \; .
\end{eqnarray}

We represent the linear transformation between the ``before'' and ``after''
combination coefficients as a $2 \times 2$ matrix $\mathbf{T}(t_1)$,
\begin{equation}
\left(\matrix{c_2 \cr d_2}\right) = \mathbf{T}(t_1) \left(\matrix{c_1 \cr 
d_1}\right) \; .
\end{equation}
Solving for its components is a straightforward exercise in linear algebra,
\begin{eqnarray}
\lefteqn{\mathbf{T}^{11}(t_1) = \Bigl(\mathbf{T}^{22}(t_1)\Bigr)^* } 
\nonumber \\
& & = {1 \over W_2} \left[ {\cal H}_{11} {\cal H}^*_{12 , x} - \left({1 - y_1
\over 1 - y_2}\right) {\cal H}_{11 , x} {\cal H}^*_{12} \right] \; , \\
\lefteqn{\mathbf{T}^{21}(t_1) = \Bigl(\mathbf{T}^{12}(t_1)\Bigr)^* } 
\nonumber \\
& & = {1 \over W_2} \left[- {\cal H}_{11} {\cal H}_{12 , x} + \left({1 - y_1 
\over 1 - y_2}\right) {\cal H}_{11 , x} {\cal H}_{12} \right] \; .
\end{eqnarray}
Here $W_2 \equiv W\Bigl(x(t_1),y_2\Bigr)$ and $W(x,y)$ is the Wronskian of our 
constant $y$ solution (\ref{consol}),
\begin{equation}
W(x,y) \equiv {\cal H} \, {\cal H}^*_{,x} - {\cal H}_{,x} \, {\cal H}^* 
= - \frac{4 i}{\pi} x^{2\mu -1} {\rm sgn}(\lambda) \; .
\end{equation}

Now consider a sequence of such transitions at times $t_1 < t_2 < t_3 < \dots$
The solution for $t > t_{n-1}$ can be expressed as the inner product, through
a product of transformation matrices $\mathbf{T}(t_k)$, between the row vector 
$({\cal H} \; , \,{\cal H}^*)$ and the column vector composed of the original 
combination coefficients,
\begin{equation}
u_n(t,k) = \Bigl( {\cal H}(x,y_n) \;\; , \;\, {\cal H}^*(x,y_n) \Bigr) 
\prod_{k = 1}^{n-1} \mathbf{T}(t_{n-k}) \left(\matrix{c_1 \cr d_1}\right) \; .
\end{equation}
We can approach continuous evolution in $y(t)$ by representing any fixed 
interval as a series of more and more transitions of ever smaller magnitude. 
In the infinitesimal limit the product of the $\mathbf{T}$ matrices 
exponentiates. 

To demonstrate exponentiation consider the transition at $t = t_1$ for $y_1 
= y(t_1) - \frac12 {\Delta y}$ and $y_2 = y(t_1) + \frac12 {\Delta y}$. The 
Wronskian and the constant $y$ solutions can all be expanded in powers of 
${\Delta y}$,
\begin{eqnarray}
\left({1 - y_1 \over 1 - y_2}\right) & = & 1 + {{\Delta y} \over 1 - y} + 
\dots \; , \\
{1 \over W_2} & = & {1 \over W} - {W_{,y} \over W^2} {{\Delta y} \over 2} +
\dots \; , \\
{\cal H}_{11} & = & {\cal H} - {\cal H}_{,y} {{\Delta y} \over 2} + 
\dots \; , \\
{\cal H}_{12} & = & {\cal H} + {\cal H}_{,y} {{\Delta y} \over 2} + 
\dots \; .
\end{eqnarray}
Expanding the coefficients of $\mathbf{T}(t_1)$ to first order gives,
\begin{eqnarray}
\mathbf{T}^{11}(t_1) & = & 1 + \zeta\Bigl(x(t_1),y(t_1) \Bigr) {\Delta y} + 
\dots \; , \\
\mathbf{T}^{21}(t_1) & = & 0 + \xi\Bigl(x(t_1),y(t_1) \Bigr) {\Delta y} + 
\dots \; , 
\end{eqnarray}
where we define,
\begin{eqnarray}
\zeta(x,y) & \equiv & {1 \over W} \left[ - {\cal H}_{,y} {\cal H}^*_{,x} + {\cal
H}_{,xy} {\cal H}^* - {{\cal H}_{,x} {\cal H}^* \over 1 - y} \right] \; , 
\label{zeta} \\
\xi(x,y) & \equiv & {1 \over W} \left[ {\cal H}_{,y} {\cal H}_{,x} - {\cal H
}_{,xy} {\cal H} + {{\cal H}_{,x} {\cal H} \over 1 - y} \right] \; . 
\label{xi}
\end{eqnarray}

Taking the infinitesimal limit gives the path-ordered product of the 
exponential of the line integral of the following matrix:
\begin{equation}
\mathbf{A}(t,k) \equiv \dot{y}(t) \left(\matrix{\zeta(x,y) & \xi^*(x,y) \cr 
\xi(x,y) & \zeta^*(x,y)} \right) \; . \label{A}
\end{equation}
We define the {\it transfer matrix} as,
\begin{eqnarray}
\mathbf{M}(t,t_i,k) & \equiv & P\left\{ \exp\left[ \int_{t_i}^t dt' 
\mathbf{A}(t',k) \right] \right\} \; , \label{M1} \\
& \equiv & \sum_{n=0}^{\infty} \int_{t_i}^t dt_1 \int_{t_i}^{t_1} dt_2 \dots 
\int_{t_i}^{t_{n-1}} dt_n \nonumber \\
& & \hspace{2cm} \times \mathbf{A}(t_1,k) \cdots \mathbf{A}(t_n,k) . \quad
\label{M2}
\end{eqnarray}
Here $t_i$ is the initial time. The full solution assumes the form,
\begin{equation}
u(t,k) = \Bigl({\cal H}(x,y) \;\; , \;\, {\cal H}^*(x,y)\Bigr) 
\mathbf{M}(t,t_i,k) \left(\matrix{c \cr d} \right) \; . \label{theans}
\end{equation}

Explicitly verifying that (\ref{theans}) obeys (\ref{diffeqn}) is facilitated
by two identities implied by the Wronskian,
\begin{eqnarray}
\zeta {\cal H} + \xi {\cal H}^* & = & - {\cal H}_{,y} \; , \\
\zeta {\cal H}_{,x} + \xi {\cal H}^*_{,x} & = & - {\cal H}_{,xy} +
{{\cal H}_{,x} \over 1 - y} \; .
\end{eqnarray}
Another very useful fact is that the time derivative of the transfer matrix
is proportional to it,
\begin{equation}
\dot{\mathbf{M}}(t,t_i,k) = \mathbf{A}(t,k) \, \mathbf{M}(t,t_i,k)
\end{equation}
Finally, we recall that time derivatives of ${\cal H}\Bigl(x(t,k),y(t)\Bigr)$
follow from the chain rule,
\begin{equation}
\dot{\cal H}(x,y) = \dot{x} \, {\cal H}_{x} + \dot{y} \, {\cal H}_{,y} \; .
\end{equation}

With these facts in hand it is straightforward to compute the first 
derivative,
\begin{eqnarray}
\dot{u}(t,k) & = & \left[ \Bigl( \dot{\cal H} \;\; , \;\, \dot{\cal H}^* \Bigr)
+ \Bigl({\cal H} \;\; , \;\, {\cal H}^*\Bigr) \mathbf{A} \right] \mathbf{M} 
\left(\matrix{c \cr d} \right) , \qquad \\
& = & \dot{x} \Bigl( {\cal H}_{,x} \;\; , \; \,{\cal H}^*_{,x} \Bigr) 
\mathbf{M} \left(\matrix{c \cr d} \right) \; .
\end{eqnarray}
The second derivative requires only a little more effort,
\begin{eqnarray}
\lefteqn{\ddot{u}(t,k) = \left[ \ddot{x} \Bigl( {\cal H}_{,x} \;\; , \;\,
{\cal H}^*_{,x} \Bigr) + \dot{x} \Bigl( \dot{\cal H}_{,x} \;\; , \;\, 
\dot{\cal H}^*_{,x} \Bigr) \right.} \nonumber \\
& & \hspace{2.5cm} \left.  + \dot{x} \Bigl( {\cal H}_{,x} \;\; , \;\, 
{\cal H}^*_{,x} \Bigr) \mathbf{A} \right] \mathbf{M} \left(\matrix{c \cr d} 
\right) , \qquad \\
& & = \Bigl( \ddot{x} \, {\cal H}_{,x} + \dot{x}^2 \, {\cal H}_{,xx} + {\dot{x} 
\, {\cal H}_{,x} \over 1 - y} \;\; , \;\, {\rm c.c.} \Bigr) \mathbf{M} \left(
\matrix{c \cr d} \right) , \qquad \\
& & = -H^2 \Bigl(x^2 + {\scriptstyle \frac{D\!-\!1}2} y - ({\scriptstyle
\frac{D\!-\!1}2})^2 \Bigr) u(t,k) \; .
\end{eqnarray}
In the last step we have substituted expressions (\ref{xdot}-\ref{xddot}) for 
$\dot{x}$ and $\ddot{x}$, and we have exploited relation (\ref{H,xx}) for 
${\cal H}_{,xx}$.

{\it 5. Evaluating the transfer matrix:} During most periods of interesting 
cosmologies one has $\vert \dot{y}/H \vert \ll 1$. Since $\dot{\mathbf{M}}$ 
vanishes with $\dot{y}$ it is reasonable to expect that the mode functions 
derive most of their time dependence from the constant deceleration solutions, 
${\cal H}(x,y)$ and ${\cal H}^*(x,y)$. This is one reason our solution is 
effective in spite of the cumbersome definition (\ref{zeta}-\ref{M2}) of
the transfer matrix. The role of $\mathbf{M}(t,t_i,k)$ is to slowly mix 
positive and negative frequency solutions as cosmological particle production 
progresses. It {\it must} be a nonlocal summation over the past.

The nonlocal character of ${\bf M}(t,t_i,k)$ does not preclude learning a 
great deal about it. We begin by recalling that ${\cal H}(x,y) \equiv x^{\mu} 
H_{\nu}^{(1)}(\lambda x)$ with $\mu = \frac12 - \frac12 \lambda$ and 
$\nu = \frac12 + (\frac{D}2 - 1) \lambda$. Now apply the chain rule 
to show,
\begin{eqnarray}
{\cal H}_{,x} & = & \mu x^{\mu-1} H_{\nu}^{(1)}(\lambda x) + \lambda x^{\mu} 
H_{\nu}^{(1)\prime}(\lambda x) \; , \\
{\cal H}_{,y} & = & \mu' \ln(x) x^{\mu} H_{\nu}^{(1)}(\lambda x) \nonumber \\
& & + \lambda' x^{\mu+1} H_{\nu}^{(1)\prime}(\lambda x) + \nu' x^{\mu} 
\partial_{\nu} H_{\nu}^{(1)}(\lambda x) \; .
\end{eqnarray}
Here a prime means differentiation with repsect to the function's argument.
Upon substitution in the definition (\ref{zeta}) for $\zeta(x,y)$ and making 
some tedious rearrangements we find,
\begin{eqnarray}
\lefteqn{\zeta = \frac12 \lambda^2 \ln(x) + \frac12 \lambda } \nonumber \\
& & - \frac{i \pi}4 \vert \lambda \vert \Bigl\{ \frac12 E_{\nu}'(\lambda x) + 
F_{\nu}( \lambda x) + (\nu - {\scriptstyle \frac12}) G_{\nu}(\lambda x) 
\Bigr\} , \qquad \end{eqnarray}
where $E_{\nu}(z)$, $F_{\nu}(z)$ and $G_{\nu}(z)$ are the following real
products of Hankel functions,
\begin{eqnarray}
E_{\nu}(z) & \equiv & z \Bigl\Vert H_{\nu}^{(1)}(z) \Bigr\Vert^2 , \\
F_{\nu}(z) & \equiv & z^2 \Biggl[ \Bigl\Vert \partial_z H_{\nu}^{(1)}(z) 
\Bigr\Vert^2 \!\!\! + \! \Bigl(1 - \frac{\nu^2}{z^2} \Bigr) \Bigl\Vert H_{\nu
}^{(1)}(z) \Bigr\Vert^2 \Biggr] , \qquad \\
G_{\nu}(z) & \equiv & z \Biggl[\partial_{\nu} H_{\nu}^{(1)}(z) \partial_z 
\Bigl(H_{\nu}^{(1)}(z)\Bigr)^* \nonumber \\
& & \hspace{2cm} - \partial_{\nu} \partial_z H_{\nu}^{(1)}(z) \Bigl(H_{\nu}^{
(1)}(z)\Bigr)^* \Biggr] .
\end{eqnarray}
It might be thought that these three functions require separate expansions
but the last two actually follow from $E_{\nu}(z)$ through the identities,
\begin{equation}
F_{\nu}'(z) = 2 E_{\nu}(z) \qquad , \qquad G_{\nu}'(z) = - {2 \nu \over z^2}
E_{\nu}(z) \; . \label{rel1}
\end{equation}

The same procedure can be used to bring $\xi(x,y)$ to a closely related form,
\begin{equation}
\xi = \frac{i \pi}4 \vert \lambda \vert \Bigl\{ \frac12 {\cal E}_{\nu}'(
\lambda x) + {\cal F}_{\nu}(\lambda x) + (\nu - {\scriptstyle \frac12}) 
{\cal G}_{\nu}(\lambda x) \Bigr\} . \qquad
\end{equation}
In this case each function is the product of the same two Hankel functions,
\begin{eqnarray}
{\cal E}_{\nu}(z) & \equiv & z \Bigl( H_{\nu}^{(1)}(z) \Bigr)^2 , \\
{\cal F}_{\nu}(z) & \equiv & z^2 \Biggl[ \Bigl( \partial_z H_{\nu}^{(1)}(z) 
\Bigr)^2 \!\!\! + \! \Bigl(1 - \frac{\nu^2}{z^2} \Bigr) \Bigl( H_{\nu}^{(1)}(z)
\Bigr)^2 \Biggr] , \qquad \\
{\cal G}_{\nu}(z) & \equiv & z \Biggl[\partial_{\nu} H_{\nu}^{(1)}(z) 
\partial_z H_{\nu}^{(1)}(z) \nonumber \\
& & \hspace{2cm} - \partial_{\nu} \partial_z H_{\nu}^{(1)}(z) H_{\nu}^{(1)}(z) 
\Biggr] .
\end{eqnarray}
These functions are not purely real, unlike those for $\zeta$. However,
they still bear the same relation to one another,
\begin{equation}
{\cal F}_{\nu}'(z) = 2 {\cal E}_{\nu}(z) \qquad , \qquad {\cal G}_{\nu}'(z) = 
- {2 \nu \over z^2} {\cal E}_{\nu}(z) \; . \label{rel2}
\end{equation}

It is straightforward to obtain explicit series expansions for $\zeta(x,y)$
and $\xi(x,y)$. One begins by expressing $E_{\nu}(z)$ and ${\cal E}_{\nu}(z)$
in terms of Bessel functions,
\begin{eqnarray}
\lefteqn{E_{\nu}(z) = {1 \over \sin^2(\nu \pi)} \Bigl\{ z J_{-\nu}^2(z) }
\nonumber \\
& & \hspace{1cm} - 2 \cos(\nu \pi) J_{-\nu}(z) J_{\nu}(z) + z J_{\nu}^2(z) 
\Bigr\} \; , \\
\lefteqn{{\cal E}_{\nu}(z) = {1 \over \sin^2(\nu \pi)} \Bigl\{ -z 
J_{-\nu}^2(z) } \nonumber \\
& & \hspace{1cm} + 2 e^{-i \nu \pi} J_{-\nu}(z) J_{\nu}(z) - e^{-2 i \nu 
\pi} z J_{\nu}^2(z) \Bigr\} \; .
\end{eqnarray}
Now exploit relations (\ref{rel1}) and (\ref{rel2}) to show,
\begin{eqnarray}
\lefteqn{\zeta = \frac12 \lambda^2 \ln(x) + \frac12 \lambda - \left[{i \pi 
(\nu - \frac12) \vert \lambda \vert \over 4 \sin^2(\nu \pi)} \right] \Bigl\{ 
b_{\nu}(\vert \lambda \vert x) \! - \! 2 \! } \nonumber \\
& & + 4 \sin^2(\nu \pi) \theta(-\lambda) \! + \! \cos(\nu \pi) 
c_{\nu}(\vert \lambda \vert x) \! + \! d_{\nu}( \vert \lambda \vert x) 
\Bigr\} , \quad \\
\lefteqn{\xi = - \left[{i \pi (\nu - \frac12) \vert \lambda \vert \over 4 
\sin^2(\nu \pi)} \right] \Bigl\{ b_{\nu}(\lambda x) - 2 } \nonumber \\
& & \hspace{2cm} + e^{-i \nu \pi} c_{\nu}(\lambda x) + e^{-2 i \nu \pi} 
d_{\nu}(\lambda x) \Bigr\} ,
\end{eqnarray}
where the various coefficient functions are,
\begin{eqnarray}
b_{\nu}(z) \!\!\! & = & \!\!\! {1 \over 2 \sqrt{\pi}} \sum_{n=1}^{\infty} 
{(-1)^n \Gamma(n \! - \! \nu \! - \! \frac12) z^{2n \! - \! 2\nu} (n \! - \!
\nu)^{- \! 1} \over \Gamma(n) \Gamma(n - \nu + 1) \Gamma(n - 2\nu + 1)} , 
\quad \label{bnu} \\
c_{\nu}(z) \!\!\! & = & \!\!\! - {4 \over \pi} \sin(\nu \pi) \Bigl[ \psi(\nu)
- 1 - \ln({\scriptstyle \frac12} z) \Bigr] \nonumber \\
& & \!\!\! - {1 \over \sqrt{\pi}} \sum_{n=1}^{\infty} {(-1)^n \Gamma(n - 
\frac12) z^{2n} n^{-1} \over \Gamma(n \! + \! \nu) \Gamma(n \! + \! 1) 
\Gamma(n \! - \! \nu \! + \! 1) } , \quad \label{cnu} \\
d_{\nu}(z) \!\!\! & = & \!\!\! {1 \over 2 \sqrt{\pi}} \sum_{n=0}^{\infty} 
{(-1)^n \Gamma(n \! + \! \nu \! - \! \frac12) z^{2n \! + \! 2\nu} (n \! + \!
\nu)^{- \! 1} \over \Gamma(n + 2 \nu) \Gamma(n + \nu + 1) \Gamma(n + 1)} , 
\quad \label{dnu}
\end{eqnarray}
Recall that $\psi(z) \equiv \Gamma'(z)/\Gamma(z)$.

It is illuminating to express the matrix $\mathbf{A}(t,k)$ (\ref{A}) in terms 
of the real and imaginary parts of $\zeta$ and $\xi$,
\begin{eqnarray}
\mathbf{A}(t,k) \!\!\! & = & \!\!\! \dot{y} {\rm Re}(\zeta) \mathbf{I} \! + \! 
\dot{y} \Bigl( \! {\rm Re}(\xi) \mathbf{\sigma}^1 \!\! + \! {\rm Im}(\xi) 
\mathbf{\sigma}^2 \!\! + \! i {\rm Im}(\zeta) \mathbf{\sigma}^3 \! \Bigr) . 
\qquad \\
\!\!\! & = & \!\!\! \frac12 {d \over dt} \ln\Bigl( \lambda x^{\lambda} a\Bigr) 
\mathbf{I} + \widetilde{\mathbf{A}}(t,k) \; .
\end{eqnarray}
The unit matrix $\mathbf{I}$ commutes with all other matrices so we can factor 
this part out of path-ordered product. Since its coefficient is a total 
derivative we can also give an explicit form for the contribution it makes to
$\mathbf{M}(t,t_i,k)$,
\begin{equation}
\mathbf{M}(t,t_i,k) = \sqrt{\frac{\lambda x^{\lambda} a}{\lambda_i 
x_i^{\lambda_i} a_i}} \; \widetilde{\mathbf{M}}(t,t_i,k) \; . \label{det}
\end{equation}
The subscript ``$i$'' in this and subsequent expressions refers to quantities 
evaluated at the initial time $t=t_i$, for example, $a_i = a(t_i)$. It is 
worth emphasizing that extracting the determinant of the transfer matrix 
{\it already} gives an improvement on the previous constant $y$ solutions. 

The residual transfer matrix $\widetilde{\mathbf{M}}(t,t_i,k)$,
\begin{equation}
\widetilde{\mathbf{M}}(t,t_i,k) \equiv P\left\{ \exp\left[ \int_{t_i}^t dt'
\widetilde{\mathbf{A}}(t',k) \right] \right\} \; ,
\end{equation}
has unit determinant and can be represented in terms of a nonnegative real
number $\rho(t,t_i,k)$ and the phases $\tau(t,t_i,k)$ and $\omega(t,t_i,k)$,
\begin{equation}
\widetilde{\mathbf{M}}(t,t_i,k) \equiv \left(\matrix{\sqrt{1 + \rho^2} e^{i 
\tau} & \rho e^{-i \omega} \cr \rho e^{i \omega} & \sqrt{1 + \rho^2} e^{-i 
\tau}} \right) \; . 
\end{equation}
We can use these parameters to express the fully normalized positive frequency
solution. First note that most of the time dependence of the full solution
resides in ${\cal H}(x,y)$ and ${\cal H}^*(x,y)$. Next note that the argument 
of the Hankel function in ${\cal H}(x,y)$ is $\lambda x$. From the time 
derivative of this quantity,
\begin{equation}
{d \over dt} \Bigl( \lambda x \Bigr) = - H x \left[ 1 - \lambda^2 
\frac{\dot{y}}{H} \right] \; ,
\end{equation}
we see that $\lambda x$ is typically a decreasing function of time. It is
therefore reasonable to choose the positive frequency solution to be initially 
pure ${\cal H}(x,y)$. With our normalization (\ref{norm}) this means that 
the full solution is,
\begin{equation}
u(t,k) = \sqrt{\pi \lambda x^{\lambda} a \over 4 k} \Bigl\{ \sqrt{1 +
\rho^2} e^{i \tau} {\cal H}(x,y) + \rho e^{i \omega} {\cal H}^*(x,y)\Bigr\} .
\label{full}
\end{equation}

{\it 6. The ultraviolet regime:} One can obtain a simple approximate form
in the ultraviolet regime of $x \gg 1$. Recall that the part of $\zeta(x,y)$ 
which appears in the residual transfer matrix is,
\begin{equation}
\widetilde{\zeta}(x,y) \! \equiv \!\! - \frac{i \pi}4 \lambda \Bigl\{ \frac12 
E_{\nu}'(\lambda x) + \! F_{\nu}(\lambda x) + (\nu - {\scriptstyle \frac12}) 
G_{\nu}(\lambda x) \Bigr\} . \quad
\end{equation}
Recall also the well-known asymptotic expansion for the Hankel function of 
the first kind,
\begin{eqnarray}
H_{\nu}^{(1)}(z) \!\!\! & = & \!\!\! \sqrt{2 \over \pi z} e^{i (z \! - \! \nu 
\frac{\pi}2 \! - \! \frac{\pi}4)} \sum_{n=0}^{\infty} {\Gamma(\nu \! + \! n 
\! + \! \frac12) \over n! \Gamma(\nu \! - \! n \! + \! \frac12)} \Bigl({i 
\over 2 i z}\Bigr)^n \! , \qquad \label{Hankel} \\
\!\!\! & \equiv & \!\!\! \sqrt{2 \over \pi z} e^{i (z \! - \! \nu \frac{\pi}2 
\! - \! \frac{\pi}4) } h_{\nu}(z) \; . \label{smallh}
\end{eqnarray}
With this to set the integration constant we can use relations (\ref{rel1})
to derive the following exact results,
\begin{eqnarray}
\frac12 E'_{\nu}(z) \! + \! F_{\nu}(z) \!\!\! & = & \!\!\! \frac{4}{\pi} z 
\! - \! 2 (\nu^2 \! - \! {\scriptstyle \frac14}) z \int_z^{\infty} \!\!\!\!\!
dz' {E_{\nu}(z') \over z^{\prime 3}} , \quad \\
G_{\nu}(z) \!\!\! & = & \!\!\! -2 + 2 \nu \int_z^{\infty} \!\!\!\!\! dz' 
{E_{\nu}(z') \over z^{\prime 2}} \; .
\end{eqnarray}
Now substitute (\ref{Hankel}) and integrate termwise to obtain an asymptotic
expansion for $\widetilde{\zeta}(x,y)$,
\begin{eqnarray}
\lefteqn{\widetilde{\zeta}(x,y) = -i \lambda^2 x + i {\scriptstyle \frac{\pi}2}
(\nu - {\scriptstyle \frac12}) \lambda } \nonumber \\
& & \! - \! {i (\nu \! - \! \frac12) \lambda \over 2 \sqrt{\pi}} \sum_{n=0}^{
\infty} {\Gamma(n \! + \! \frac12) \Gamma(\nu \! + \! n \! + \! \frac12) \over 
(n \!  + \! 1)! \Gamma(\nu \! - \! n \! - \! \frac12)} {(\lambda x)^{\! - \! 
2n\! - \! 1} \over (2n + 1)} \; . \quad 
\end{eqnarray}
It is simplest to express the asymptotic expansion for $\xi(x,y)$ in terms of
the function $h_{\nu}(z)$ defined in (\ref{smallh}),
\begin{eqnarray}
\lefteqn{\xi(x,y) = \frac{i}2 \lambda e^{2i (\lambda x - \nu \frac{\pi}2 -
\frac{\pi}4)} \Bigl\{ i z [h_{\nu}^2(z)]' - (\nu^2 - {\scriptstyle \frac14})
{h^2_{\nu}(z) \over z} } \nonumber \\
& & + z [h^{\prime}_{\nu}(z)]^2 \! + \! \partial_{\nu} h_{\nu}(z) h_{\nu}'(z) 
\! - \!  \partial_{\nu} h_{\nu}'(z) h_{\nu}(z) \Bigr\}_{\! z = \lambda x} 
\!\!\!\!\!\! . \qquad
\end{eqnarray}

From the expansion of $h_{\nu}(z)$ we see that $\xi(x,y)$ is negligible in
the ultraviolet,
\begin{equation}
\xi(x,y) = \frac{i}2 \lambda e^{2i (\lambda x - \nu \frac{\pi}2 - \frac{\pi}4)}
\Bigl\{ - {i (\nu - \frac12) \over 2 (\lambda x)^2} + O\Bigl( (\lambda x)^{-3}
\Bigr) \Bigr\} \; .
\end{equation}
Ignoring $\xi(x,y)$ defines the ultraviolet regime,
\begin{equation}
\widetilde{\mathbf{A}}_{\rm uv}(t,k) \equiv \dot{y} \, \widetilde{\zeta}(x,y)
\left(\matrix{1 & 0 \cr 0 & -1}\right) \; .
\end{equation}
In the notation of (\ref{full}) the ultraviolet regime is characterized by,
\begin{eqnarray}
\rho_{\rm uv}(t,t_i,k) & = & 0 \; , \label{UV1} \\
\tau_{\rm uv}(t,t_i,k) & = & \int_{t_i}^t dt' \dot{y} \, \widetilde{\zeta}(x,y)
\; .  \label{UV2}
\end{eqnarray}
Since $\rho_{\rm uv} = 0$ the value of $\omega_{\rm uv}$ is irrelevant.

In the extreme ultraviolet limit we need only keep the first two terms from
the expansion of $\widetilde{\zeta}(x,y)$,
\begin{eqnarray}
\tau_{\rm uv}(t,t_i,k) & \longrightarrow & \int_{t_i}^t dt' \dot{y} \Bigl[ - 
\lambda^2 x + {\scriptstyle \frac{\pi}2} (\nu - {\scriptstyle \frac12}) 
\lambda \Bigr] \; ,\\
& = & \Bigl[- \lambda x + {\scriptstyle \frac{\pi}2} (\nu - {\scriptstyle
\frac12}) \Bigr] \Bigl\vert_{t_i}^t - k \! \int_{t_i}^t \! {dt' \over a(t')} . 
\quad
\end{eqnarray}
The integral proportional to $-k$ on the last line is just the conformal time
interval $\eta - \eta_i$. Substituting as well the asymptotic expansion of the 
Hankel function in ${\cal H}(x,y)$ gives the following leading ultraviolet 
result,
\begin{equation}
u_{\rm uv}(t,k) \longrightarrow \sqrt{\frac{a(t)}{2k}} \, 
e^{-i k (\eta \!-\! \eta_i) + i \chi_i} \; ,
\end{equation}
where $\chi(t,k) \equiv \lambda(t) \, x(t,k) \!-\! [\nu(t) \!+\! \frac12] 
\frac{\pi}2$. Recovering the correct correspondence limit is an important 
check on the consistency of the formalism.

{\it 7. The infrared regime:} For phenomenology it is the infrared regime 
($x \ll 1$) that is more useful. In this case it is best to decompose the 
Hankel functions into Bessel functions of positive and negative order,
\begin{equation}
{\cal H}(x,y) = i {\rm csc}(\nu \pi) x^{\mu} \Bigl[e^{-i \nu \pi} J_{\nu}(
\lambda x) - J_{\!-\!\nu}( \lambda x) \Bigr] .
\end{equation}
The conjugate solution takes slightly different forms for $\lambda$ positive
(inflation) and negative (subluminal expansion),
\begin{eqnarray}
\lefteqn{{\cal H}^*\!(x,y) = i {\rm csc}(\nu \pi) x^{\mu} \Bigl[- e^{i \nu \pi 
{\rm sgn}(\lambda)} J_{\nu}(\lambda x) } \nonumber \\
& & \hspace{3.5cm} + e^{2 i \nu \pi \theta(-\lambda)} J_{\!-\!\nu}( \lambda x) 
\Bigr] . \quad
\end{eqnarray}
We seek a matrix ${\bf O}(t)$ that effects the change of basis from ${\cal H}$
and ${\cal H}^*$ to solutions based upon $J_{\!-\!\nu}$ and $J_{\nu}$,
\begin{eqnarray}
\lefteqn{\frac1{\sqrt{2}} \Bigl( {\cal H}(x,y) \, , \, {\cal H}^*\!(x,y) \Bigr) 
{\bf O}(t) } \nonumber \\
& & = x^{\mu} \Bigl( -i {\rm csc}(\nu \pi) J_{\!-\!\nu}(\lambda x) \, , \,
J_{\nu}(\lambda x) \Bigr) .
\end{eqnarray}
For positive $\lambda$ (inflation) the required matrix and its inverse are,
\begin{eqnarray}
{\bf O}(t) & = & \frac1{\sqrt{2}} \left(\matrix{1 \!-\! i {\rm cot}(\nu \pi) 
& 1 \cr -1 \!-\! i {\rm cot}(\nu \pi) & 1}\right) \; , \\
{\bf O}^{\!-\!1} & = & \frac1{\sqrt{2}} \left(\matrix{1 & -1 \cr
1 \!+\! i {\rm cot}(\nu \pi) & 1 \!-\! i {\rm cot}(\nu \pi)}\right) \; .
\end{eqnarray}
For $\lambda$ negative (subluminal expansion) we have,
\begin{eqnarray}
{\bf O}(t) & = & \frac1{\sqrt{2}} \left(\matrix{1 \!+\! i {\rm cot}(\nu \pi) 
& -e^{2i \nu \pi} \cr 1 \!+\! i {\rm cot}(\nu \pi) & -1}\right) \; , \\
{\bf O}^{\!-\!1} & = & \frac1{\sqrt{2}} \left(\matrix{1 & e^{2 i \nu \pi} \cr
1 \!+\! i {\rm cot}(\nu \pi) & -1 \!-\! i {\rm cot}(\nu \pi)}\right) \; .
\end{eqnarray}

In the new basis this matrix is,
\begin{eqnarray}
{\cal M}(t_2,t_1,k) & \equiv & {\bf O}^{\!-\!1}\!(t_2) \widetilde{\bf 
M}(t_2,t_1,k) {\bf O}(t_1,k) \; , \\
& \equiv & P\left\{ \exp\left[ \int_{t_1}^{t_2} dt {\cal A}(t,k)\right] 
\right\} \; .
\end{eqnarray}
The new exponent matrix is related to the old as follows,
\begin{equation}
{\cal A}(t,k) = \dot{\bf O}^{\!-\!1}\!(t) {\bf O}(t) + {\bf O}^{\!-\!1}\!(t)
\widetilde{\bf A}(t,k) {\bf O}(t) \; .
\end{equation}
This is just an exercise in matrix multiplication. For either sign of 
$\lambda$ the result is,
\begin{eqnarray}
{\cal A} & \!\! = \!\! & \!\frac{\pi}4 \dot{\nu} \left(\matrix{ \!\! {\rm 
csc}(\nu \pi) c_{\nu}(\lambda x) \!\! & \!\! - 2 i d_{\nu}(\lambda x) \!\! \cr 
\!\! -2 i {\rm csc}^2\!(\nu \pi) b_{\nu}(\lambda x) \!\! & \!\! -{\rm csc}(\nu 
\pi) c_{\nu}(\lambda x) \!\!} \right) , \qquad \\
& \!\! \equiv \!\! & \left(\matrix{\gamma(t,k) & -i \delta(t,k) \cr -i 
\beta(t,k) & -\gamma(t,k)} \right) \; .
\end{eqnarray}

For small $x$ the coefficients of ${\cal A}$ go like,
\begin{equation}
\beta \sim x^{2\!-\!2\nu} \; , \; \gamma \sim \ln(x) \; , \; \delta \sim
x^{2\nu} \; .
\end{equation}
Although $\beta(t,k)$ makes the dominant contribution to the residual
transfer matrix for small $x$, it does not make the dominant contribution 
to the mode function. This is because the only way a factor of $\beta$ can
multiply the $J_{\!-\!\nu}$, rather than the $J_{\nu}$, is if it accompanies
a factor of $\delta$. To see this, first consider the basis functions times
a single factor of ${\cal A}$,
\begin{eqnarray}
\lefteqn{ \Bigl(-i {\rm csc}(\nu \pi) J_{\!-\!\nu} \, , \, J_{\nu} \Bigr)
{\cal A} } \nonumber \\
& & \!\!=\! \Bigl(\!-i {\rm csc}(\nu \pi) \gamma J_{\!-\!\nu} \!\!-\! i \beta 
J_{\nu} \, , \!-{\rm csc}(\nu \pi) \delta J_{\!-\!\nu} \!\!-\! \gamma J_{\nu} 
\Bigr) . \quad
\end{eqnarray}
The small $x$ behavior of $\gamma J_{\!-\!\nu}$ is $\sim \ln(x) x^{\!-\!\nu}$, 
whereas $\beta J_{\nu} \sim x^{2\!-\!\nu}$ is down by $x^2$. Now consider 
two factors of ${\cal A}$ for any times during which $x \ll 1$,
\begin{equation}
{\cal A}(t_2,k) {\cal A}(t_1,k) \!=\! \left(\matrix{\!\! \gamma_2 
\gamma_1 \!-\! \delta_2 \beta_1 \!\!\! & \!\!\! -i [\gamma_2 \delta_1 \!-\! 
\delta_2 \gamma_1] \cr \!\! -i [\beta_2 \gamma_1 \!-\! \gamma_2 \beta_1] 
\!\!\! & \!\!\! \gamma_2 \gamma_1 \!-\! \beta_2 \delta_1 \!\!} \right) .
\end{equation}
Since $\beta \delta \sim x^2$, these terms are negligible with respect to
$\gamma^2 \sim \ln^2(x)$.

To the same proportional error of order $x^2 \ll 1$ we can easily obtain an
explicit expression for the $n$-fold product,
\begin{eqnarray}
\lefteqn{{\cal A}(t_n,k) \cdots {\cal A}(t_1,k) \longrightarrow} \nonumber
\\ & & \!\!\! \left(\matrix{\!\! \gamma_n \gamma_{n\!-\!1} \cdots \gamma_1 & 
\! -i [\gamma_n \cdots \gamma_2 \delta_1 \!-\! \dots] \cr \!\! -i 
[\beta_n \gamma_{n\!-\!1} \cdots \gamma_1 \!-\! \dots]  & \! (-1)^n 
\gamma_n \gamma_{n\!-\!1} \cdots \gamma_1 \!\!} \right) . \qquad
\end{eqnarray}
In this expression the off diagonal elements are,
\begin{eqnarray}
\lefteqn{-i [\gamma_n \cdots \gamma_2 \delta_1 \!-\! \dots]  \equiv -i \Bigl[
\gamma_n \cdots \gamma_2 \delta_1 } \nonumber \\
& & \hspace{.75cm} \!-\! \gamma_n \cdots \gamma_3 \delta_2 \gamma_1 \!+\! \dots 
\!-\!(-1)^n \delta_n \gamma_{n\!-\!1} \cdots \gamma_1\Bigr] , \qquad \\
\lefteqn{-i [\beta_n \gamma_{n\!-\!1} \cdots \gamma_1 \!-\! \dots] \equiv -i 
\Bigl[\beta_n \gamma_{n\!-\!1} \cdots \gamma_1 } \nonumber \\
& & \hspace{.75cm} \!-\! \gamma_n \delta_{n\!- \!1} \gamma_{n\!-\!2} \cdots 
\gamma_1 \!+\! \dots \!-\!(-1)^n \gamma_n \cdots \gamma_2 \beta_1 \Bigr] . 
\qquad
\end{eqnarray}
This form can be exponentiated,
\begin{equation}
{\cal M}_{\rm ir}(t_2,t_1,k) \longrightarrow \left(\matrix{\Gamma(t_2,t_1,k) 
&-i \Delta(t_2,t_1,k) \cr -i {\rm B}(t_2,t_1,k) & \Gamma^{\!-\!1}\!(t_2,t_1,k)} 
\right) , \label{expform}
\end{equation}
where the various components are,
\begin{eqnarray}
\Gamma(t_2,t_1,k) & \!=\! & \exp\left[ \int_{t_1}^{t_2} \!\! dt \gamma(t,k) 
\right] \; , \label{Gamma} \\
{\rm B}(t_2,t_1,k) & \!=\! & \int_{t_1}^{t_2} \!\! dt \Gamma^{-\!1}\!(t_2,
t,k) \beta(t,k) \Gamma(t,t_1,k) , \quad \label{Beta} \\
\Delta(t_2,t_1,k) & \!=\! & \int_{t_1}^{t_2} \!\! dt \Gamma(t_2,t,k) 
\delta(t,k) \Gamma^{-\!1}\!(t,t_1,k) . \quad \label{Delta}
\end{eqnarray}

Because (\ref{expform}) is only valid up to factors of order $x^2$ it is 
superfluous to evaluate the components using anything but the leading
small $x$ forms for $\beta$, $\gamma$ and $\delta$,
\begin{eqnarray}
\beta(t,k) & \! = \! & \frac1{2\pi} \frac{\dot{\nu}}{\nu-\frac12} 
\frac{\Gamma^2(\nu-1)}{(\frac12 \lambda x)^{2 \nu - 2}} + O(x^{4-2\nu}) \; , \\
\gamma(t,k) & \! = \! & \dot{\nu} \Bigl[\ln(\frac12 \lambda x) + 1 - \psi(\nu)
\Bigr] + O(x^2) \; , \label{gam0} \\
\delta(t,k) & \! = \! & \frac{\pi}2 \frac{\dot{\nu}}{\nu-\frac12} 
\frac{(\frac12 \lambda x)^{2 \nu}}{\Gamma^2(\nu-1)} + O(x^{2\nu+2}) \; .
\end{eqnarray}
The leading term in (\ref{gam0}) is a total derivative,
\begin{equation}
\gamma(t,k) = \frac{d}{dt} \ln\left[\frac{a^{\frac{D}2 - 1} (\frac12 \lambda x
)^{\nu - \frac12}}{\Gamma(\nu)}\right] + O(x^2) \; , 
\end{equation}
so we can obtain explicit expressions for $\Gamma$, ${\rm B}$ and $\Delta$,
up to correction factors of order $x^2$,
\begin{eqnarray}
\Gamma(t_2,t_1,k) & \!\! \rightarrow \!\! & \frac{a_2^{\frac{D}2 \!-\! 1} 
(\frac12 \lambda_2 x_2)^{\nu_2 \!-\! \frac12}}{\Gamma(\nu_2)} \frac{\Gamma(
\nu_1)}{a_1^{\frac{D}2 \!-\! 1} (\frac12 \lambda_1 x_1)^{\nu_1 \!-\!\frac12}} 
, \qquad \\
{\rm B}(t_2,t_1,k) & \!\! \rightarrow \!\! & \frac{\Gamma(\nu_2)}{a_2^{\frac{D}2
\!-\! 1} (\frac12 \lambda_2 x_2)^{\nu_2 \!-\! \frac12}} \frac{\Gamma(\nu_1)}{
a_1^{\frac{D}2 \!-\! 1} (\frac12 \lambda_1 x_1)^{\nu_1 \!-\! \frac12}} 
\nonumber \\
& & \times \frac1{2\pi (D\!-\!2)} \int_{t_1}^{t_2} \!\! dt \frac{\dot{\nu}}{
(\nu \!-\!1)^2} x a^{D-2} , \\
\Delta(t_2,t_1,k) & \!\! \rightarrow \!\! & \frac{a_2^{\frac{D}2 \!-\! 1} 
(\frac12 \lambda_2 x_2)^{\nu_2 \!-\! \frac12}}{\Gamma(\nu_2)} \frac{
a_1^{\frac{D}2 \!-\! 1} (\frac12 \lambda_1 x_1)^{\nu_1 \!-\! \frac12}}{
\Gamma(\nu_1)} \nonumber \\
& & \times \frac{\pi^2}{2 \pi (D\!-\!2)} \int_{t_1}^{t_2} \!\! dt 
\frac{\dot{\nu}}{\nu^2} \frac{x}{a^{D-2}} .
\end{eqnarray}

It is useful to compare the elements of the same column of the infrared
transfer matrix (\ref{expform}). The ratio of the 21 and 11 components is,
\begin{eqnarray}
\lefteqn{\frac{-i B(t_2,t_1,k)}{\Gamma(t_2,t_1,k)} \longrightarrow \frac{-i
\Gamma^2(\nu_2)}{2 \pi (D\!-\!2)} \, x_2 \Bigl(\frac12 \lambda_2 x_2\Bigr)^{1 
\!-\! 2\nu_2} } \nonumber \\
& & \hspace{3cm} \times \int_{t_1}^{t_2} \!\! dt \frac{\dot{\nu}}{(\nu 
\!-\!1)^2} \frac{H_2}{H} \Bigl(\frac{a}{a_2}\Bigr)^{\!D\!-3} \!\!\! . \qquad
\end{eqnarray}
In the infrared regime we know $x_2 \ll 1$. The expansion of the universe
implies $a(t)/a_2 < 1$, and the weak energy condition tells us $H_2/H(t) < 
1$. During inflation the parameter $\nu(t)$ is positive ($\nu \ge \frac{D-1}2$)
so the factor $x_2^{2\!-\!2\nu_2}$ can win out over the other terms and make 
$-i B(t_2,t_1,k)$ dominate $\Gamma(t_2,t_1,k)$. After inflation the parameter 
$\nu(t)$ changes sign and $-iB(t_2,t_1,k)$ is insignificant compared with 
$\Gamma(t_2,t_1,k)$. (The only exception is when $\nu_2$ passes through a 
negative integer, at which point $\Gamma(\nu_2)$ diverges and the infrared 
basis becomes degenerate.) The ratio of the 12 and 22 components is,
\begin{equation}
\frac{-i\Delta(t_2,t_1,k)}{\Gamma^{\!-\!1}\!(t_2,t_1,k)} \!\rightarrow \!
\frac{-i \pi x_2 (\frac12 \lambda_2 x_2)^{2\nu_2 \!-\! 1}}{2 (D\!-\!2) 
\Gamma^2(\nu_2)} \!\! \int_{t_1}^{t_2} \!\!\!\! dt \frac{\dot{\nu}}{\nu^2} 
\frac{H_2}{H} \Bigl(\!\frac{a_2}{a}\! \Bigr)^{\!D-\!1} \!\!\! . 
\end{equation}
Some of these factors are less than one, others are greater, and the 
magnitude of the ratio is not clear.

Recall the series expansion of the Bessel function,
\begin{equation}
J_{\nu}(z) = \sum_{n=0}^{\infty} \frac{(-1)^n (\frac12 z)^{2n + \nu}}{n! 
\Gamma(n +1 + \nu)} \; . \label{Besser}
\end{equation}
From this we see that the linearly independent solutions appropriate to
the far infrared consist of a ``big'' mode function $u^-(t,t_1,k)$,
based on $J_{-\nu}(\lambda x)$, and a ``small'' mode function $u^+(t,t_1,k)$, 
based on $J_{\nu}(\lambda x)$,
\begin{eqnarray}
u^-(t,t_1,k) &\! \equiv \!& \sqrt{\frac{\pi a}{k} \, \frac{\lambda x}2}
\Bigl(\!-i {\rm csc}(\nu \pi) J_{\!-\!\nu}(\lambda x) \, , \nonumber \qquad \\
& & \hspace{1cm} J_{\nu}(\lambda x) \Bigr) {\cal M}(t,t_1,k) 
\left(\matrix{1 \cr 0}\right) , \quad \\
u^+(t,t_1,k) &\! \equiv &\! \sqrt{\frac{\pi a}{k} \, \frac{\lambda x}2}
\Bigl(\!-i {\rm csc}(\nu \pi) J_{\!-\!\nu}(\lambda x) \, , \nonumber \qquad \\
& & \hspace{1cm} J_{\nu}(\lambda x) \Bigr) {\cal M}(t,t_1,k) \left(\matrix{0 
\cr 1}\right) , \quad
\end{eqnarray}
It is straightforward to verify that these solutions obey,
\begin{equation}
u^-(t,t_1,k) \dot{u}^+(t,t_1,k) - \dot{u}^-(t,t_1,k) u^+(t,t_1,k) = i \; . 
\label{irwron}
\end{equation}

If $x \ll 1$ between $t_1$ and $t$, we need only keep the $n=0$ terms of
the Bessel functions (\ref{Besser}). We can also employ the infrared 
simplifications (\ref{expform}-\ref{Delta}) of the transfer matrix. Even 
though $B(t,t_1,k)$ can be larger than $\Gamma(t,t_1,k)$, its contribution 
to $u^-(t,t_1,k)$ is still insignificant because it multiplies the $J_{\nu}$,
\begin{eqnarray}
u^-(t,t_1,k) & \!\!\! \longrightarrow  \!\!\! & -i \sqrt{\frac{\pi a}{k}
\, \frac{\lambda x}2} \Bigl[ {\rm csc}(\nu \pi) J_{\!-\!\nu}(\lambda x)
\Gamma(t,t_1,k) \quad \nonumber \\
& & \hspace{2cm} + J_{\nu}(\lambda x) B(t,t_1,k) \Bigr] , \qquad \\
& \!\!\! \longrightarrow \!\!\! & \frac{-i a^{\!\frac{D-1}{2}}}{\sqrt{\pi k}} 
\; \frac{\Gamma(\nu_1)}{a_1^{\frac{D}2\!-\!1} (\frac12 \lambda_1 x_1)^{\nu_1 
\!-\! \frac12}} . \label{ubig}
\end{eqnarray}
Most of our limiting form (\ref{ubig}) consists of constants; its time 
dependence derives entirely from the factor of $a^{\frac{D-1}2}$. This 
corresponds to a constant scalar field, and we see from (\ref{phieqn}) that 
$\widetilde{\varphi}(t,\vec{k}) = {\rm constant}$ is indeed a solution for the 
infrared limit of $\vec{k} = 0$.

Even in the infrared limit both terms contribute to $u^+(t,t_1,k)$,
\begin{eqnarray}
\!\!\lefteqn{u^+(t,t_1,k) \! \longrightarrow \! \sqrt{\frac{\pi a}{k} \, 
\frac{\lambda x}2} \Bigl[\!-\! {\rm csc}(\nu \pi) J_{\!-\!\nu}(\lambda x) 
\Delta(t,t_1,k) } \qquad \nonumber \\
& & \hspace{3.5cm} + J_{\nu}(\lambda x) \Gamma^{-\!1}\!(t,t_1,k) \Bigr] ,
\qquad \\
& & \! \longrightarrow \! \sqrt{\pi k} \, a^{\frac{D-1}{2}} \; \frac{a_1^{
\frac{D}2\!-\!1} (\frac12 \lambda_1 x_1)^{\nu_1 \!-\! \frac12}}{\Gamma(\nu_1)} 
\nonumber \\
& & \times \Bigl[\frac{-1}{2D \! -\! 4} \int_{t_1}^t \!\! dt' \frac{\dot{\nu}
}{\nu^2 H a^{D-1}} + \frac{\lambda}{2 \nu H a^{D-1}} \Bigr] . \qquad 
\label{small1}
\end{eqnarray}
It is straightforward to partially integrate once,
\begin{equation}
\int_{t_1}^t \!\! dt' \frac{\dot{\nu}}{\nu^2 H a^{D-1}} = - \frac1{\nu H 
a^{D-1}} \Bigl\vert_{t_1}^t - \int_{t_1}^t \!\! dt' \frac{D-1-y}{\nu a^{D-1}}
\; .
\end{equation}
From the identities,
\begin{eqnarray}
\frac{D-1-y}{\nu} & = & 2 - 2 y = 2 - \frac{d}{dt} \Bigl(\frac{2}{H} \Bigr)
\; , \\
-\frac1{\nu} + 2 & = & (D-2) \frac{\lambda}{\nu} \; ,
\end{eqnarray}
we see that a second partial integration brings the integral to the form,
\begin{equation}
\frac{-1}{2D\!-\!4} \int_{t_1}^t \!\! dt' \frac{\dot{\nu}}{\nu^2 H a^{D-1}} = 
\frac{-\lambda}{2 \nu H a^{D-1}} \Bigl\vert_{t_1}^t - \int_{t_1}^t \frac{dt'}{ 
a^{D-1}} .
\end{equation}
Substitution in (\ref{small1}) gives the infrared limit for the small mode
function,
\begin{eqnarray}
\lefteqn{u^+(t,t_1,k) \! \longrightarrow \! \sqrt{\pi k} \, 
a^{\frac{D-1}{2}} \; \frac{a_1^{\frac{D}2\!-\!1} (\frac12 \lambda_1 x_1)^{
\nu_1 \!-\! \frac12}}{\Gamma(\nu_1)} } \nonumber \\
& & \hspace{2cm} \times \Bigl[\frac{\lambda_1}{2 \nu_1 H_1 a_1^{D-1}} - 
\int_{t_1}^t \frac{dt'}{a^{D-1}} \Bigr] . \label{usmall}
\end{eqnarray}
Note that the infrared limits (\ref{ubig},\ref{usmall}) of our solutions 
retain their canonical normalization (\ref{irwron}).

The two independent solutions of the scalar field equation (\ref{phieqn}) in 
the infrared limit of $\vec{k} = 0$
are,
\begin{equation}
\widetilde{\varphi}(t,\vec{0}) \sim 1 \quad , \quad \widetilde{\varphi}(t,
\vec{0}) \sim \int_{t_1}^t \frac{dt'}{a^{D-1}} \; . \label{2sols}
\end{equation}
We have seen that $u^-(t,t_1,k)$ becomes proportional to the constant
solution, but the second solution in (\ref{2sols}) also contains a constant 
term from its dependence upon the lower limit of integration. Since the 
integrand {\it falls} as the universe expands, care must be taken to isolate 
the small, time-dependent term from the potentially much larger constant. This 
is something our formalism does automatically. For if $y(t)$ is constant for 
$t \geq t_1$ we can solve for the Hubble parameter and the scale factor,
\begin{eqnarray}
H(t)\Bigl\vert_{y = y_1} & = & \frac{H_1}{1 + y_1 H_1 (t - t_1)} \; , \\
a(t)\Bigl\vert_{y = y_1} & = & a_1 \Bigl[1 + y_1 H_1 (t - t_1) \Bigr]^{
\frac1{y_1}} \; .
\end{eqnarray}
In this case the integral can be explicitly evaluated,
\begin{equation}
\int_{t_1}^t \frac{dt'}{a^{D-1}} \Bigl\vert_{y=y_1} = -\frac{\lambda_1}{
2 \nu_1 H a^{D-1}} + \frac{\lambda_1}{2 \nu_1 H_1 a_1^{D-1}} .
\end{equation}
Hence the constant on the second line of our infrared form (\ref{usmall})
for $u^+(t,t_1,k)$ cancels the large constant from the lower limit,
up to very small terms coming from the weak time dependence of $y(t)$. So our
second solution is indeed dominated by the small, time-dependent, lower
limit term.

It remains only to express the original mode solution (\ref{full}) in the 
infrared basis,
\begin{eqnarray}
\lefteqn{u(t,k) = \frac1{\sqrt{2}} \Bigl(u^-(t,t_1,k) \, , \, u^+(t,t_1,k) 
\Bigr) } \nonumber \\
& & \hspace{2cm} \times {\cal M}(t_1,t_i,k) \left(\matrix{1 \cr 1 \!+\! i 
{\rm cot}(\nu_i \pi)} \right) . \qquad
\end{eqnarray}
It is reasonable for $t_1$ to be the time of first horizon crossing, that is,
$x(t_1,k) = 1$. Even though $x(t,k)$ is just becoming small at this time,
the expansion is so rapid during inflation that we have $x \ll 1$ almost
immediately afterwards. Because the transfer matrix up to this time is of
order one, only $u^-(t,t_1,k)$ contributes significantly,
\begin{eqnarray}
\lefteqn{u_{\rm ir}(t,k) \! \longrightarrow \! \frac{-i a^{\frac{D-1}2}}{
\sqrt{2k}} \frac{\Gamma(1 \!-\!\nu) J_{\!-\nu}(\lambda x)}{(\frac12 
\lambda x)^{-\nu}} } \nonumber \\
& & \hspace{3cm} \times \Bigl(\frac{H_1}{k}\Bigr)^{\!\frac{D}2 \!-\! 1} 
{\cal C}_1(k) {\cal C}_i(k) .  \label{IRu}
\end{eqnarray}

We have expressed $u_{\rm ir}$ with the standard normalization times two
correction factors. The correction that depends upon the system at first
horizon crossing is,
\begin{equation}
{\cal C}_1(k) \equiv \frac{\frac1{\sqrt{\pi}} \Gamma(\nu_1)}{(\frac12 
\lambda_1)^{\nu_1 \!-\! \frac12}} \; . \label{C1}
\end{equation}
The other correction factor depends upon evolution up to this point,
\begin{equation}
{\cal C}_i(k) \equiv {\cal M}^{11}(t_1,t_i,k) + \frac{i e^{-i \nu_i \pi}}{
\sin(\nu_i \pi)} {\cal M}^{12}(t_1,t_i,k) . \label{Ci}
\end{equation}
 
Further corrections to (\ref{IRu}) are down by a factor $x^2(t',k)$ for some 
$t'$ between first and second horizon crossings. (Second horizon crossing is 
the time $t_2$, after inflation, at which $x(t_2,k) =1$.) Before second 
horizon crossing we need retain only the leading term in the series expansion 
of the Bessel function (\ref{Besser}), but the form (\ref{IRu}) should remain 
valid to arbitrarily late times provided that $y(t)$ does not experience 
violent evolution at late times. This is a very safe assumption given what we 
know about late time cosmology. 

{\it 8. Summary and discussion:} This has been a long and technical series
of derivations, and it was not clear at the beginning what would be the most
economical form in which to express the final answer. Now that we are done,
it is well to summarize the important results in a simple way. It is also
desirable to replace the parameter $y(t) \equiv -\dot{H}/H^2$, and the 
quantities $\lambda$ and $\mu$ which are derived from it, with the familiar
deceleration parameter, $q(t) = -1 + y(t)$. The only exception we make is
the parameter $\nu = \frac12 \!-\! (\frac{D}2 \!-\! 1) \frac1{q}$, which 
appears in too many indices to be conveniently expunged.

Our mode solution is specified in terms of parameters constructed from the 
scale factor and its derivatives,
\begin{equation}
x(t,k) \equiv \frac{k}{a H} \; , \; q(t) \equiv -1 \!-\! 
\frac{\dot{H}}{H^2} \; {\rm and} \; \nu(t) \equiv \frac12 \!-\!
\frac{\frac{D}2 \!-\! 1}{q} .
\end{equation}
The solution to equation (\ref{diffeqn}) which is positive frequency at the 
initial time $t = t_i$ takes the form of a row vector of Bessel functions
multiplied into a transfer matrix,
\begin{eqnarray}
\lefteqn{u(t,k) \! \equiv \! \sqrt{\frac{\pi a}{2k}} \Bigl(-\frac{x}{2q}
\Bigr)^{\frac12} \Bigl(\!-i {\rm csc}(\nu \pi) J_{\!-\!\nu}({\scriptstyle 
\!-\! \frac{x}{q}}) \, , \, J_{\nu}({\scriptstyle \!-\! \frac{x}{q}}) 
\Bigr) } \nonumber \\
& & \hspace{2.5cm} \times {\cal M}(t,t_i,k) \left(\matrix{1 \cr 1 \!+\! i
{\rm cot}[\nu(t_i) \pi] }\right) . \qquad \label{genu}
\end{eqnarray}
The transfer matrix ${\cal M}(t,t_i,k)$ is the time-ordered product of the 
exponential of a line integral,
\begin{eqnarray}
\lefteqn{{\cal M}(t,t_i,k) \equiv P\left\{ \exp\left[ \int_{t_i}^t dt' 
{\cal A}(t',k) \right] \right\} \; ,} \\
& & \!\!\equiv \sum_{n=0}^{\infty} \int_{t_i}^t \!\! dt_1 \!\! \int_{t_i}^{t_1} 
\!\!\! dt_2 \dots \!\int_{t_i}^{t_{n-1}} \!\!\!\!\!\!\!\!\! dt_n 
{\cal A}(t_1,k) \cdots {\cal A}(t_n,k) . \quad
\end{eqnarray}
The exponent matrix ${\cal A}(t,k)$ vanishes whenever $q(t)$ is constant. It 
has the form,
\begin{equation}
{\cal A}(t,k) \! = \! \!\frac{\pi}4 \dot{\nu} \left(\matrix{ \!\! {\rm 
csc}(\nu \pi) c_{\nu}({\scriptstyle \!-\! \frac{x}{q}}) \!\! & \!\! - 2 i 
d_{\nu}({\scriptstyle \!-\! \frac{x}{q}}) \!\! \cr \!\! -2 i {\rm csc}^2\!(\nu 
\pi) b_{\nu}({\scriptstyle \!-\! \frac{x}{q}}) \!\! & \!\! -{\rm csc}(\nu \pi) 
c_{\nu}({\scriptstyle \!-\! \frac{x}{q}}) \!\!} \right) ,
\end{equation}
where the various coefficient functions are,
\begin{eqnarray}
b_{\nu}(z) \!\!\! & = & \!\!\! {1 \over 2 \sqrt{\pi}} \sum_{n=1}^{\infty} 
{(-1)^n \Gamma(n \! - \! \nu \! - \! \frac12) z^{2n \! - \! 2\nu} (n \! - \!
\nu)^{- \! 1} \over \Gamma(n) \Gamma(n \!-\! \nu \!+\! 1) \Gamma(n \!-\! 2\nu 
\!+\! 1)} , \quad \\
c_{\nu}(z) \!\!\! & = & \!\!\! - {4 \over \pi} \sin(\nu \pi) \Bigl[ \psi(\nu)
- 1 - \ln({\scriptstyle \frac12} z) \Bigr] \nonumber \\
& & \quad - {1 \over \sqrt{\pi}} \sum_{n=1}^{\infty} {(-1)^n \Gamma(n \!-\! 
\frac12) z^{2n} n^{-1} \over \Gamma(n \! + \! \nu) \Gamma(n \! + \! 1) 
\Gamma(n \! - \! \nu \! + \! 1) } , \quad \\
d_{\nu}(z) \!\!\! & = & \!\!\! {1 \over 2 \sqrt{\pi}} \sum_{n=0}^{\infty} 
{(-1)^n \Gamma(n \! + \! \nu \! - \! \frac12) z^{2n \! + \! 2\nu} (n \! + \!
\nu)^{- \! 1} \over \Gamma(n \!+\! 2 \nu) \Gamma(n \!+\! \nu \!+\! 1) 
\Gamma(n \!+\! 1)} , 
\quad
\end{eqnarray}

In the ultraviolet ($x(t,k) \gg 1$) our general solution (\ref{genu}) 
approaches,
\begin{equation}
u_{\rm uv}(t,k) \longrightarrow \sqrt{\frac{a(t)}{2k}} \, 
e^{-i k (\eta \!-\! \eta_i) + i \chi_i} \; , \label{uUV}
\end{equation}
were $\chi(t,k) \equiv \lambda(t) \, x(t,k) \!-\! [\nu(t) \!+\! \frac12] 
\frac{\pi}2$. This is just the flat space solution with a trivial redshift
factor. Corrections to are of order $1/x$ times (\ref{uUV}). 

It is common to assume all modes are originally subhorizon ($x(t_i,k) > 1$). 
During inflation $x(t,k)$ falls, typically exponentially. This carries more 
and more modes into the infrared regime of $x(t,k) \ll 1$. First horizon
crossing is the time, $t_1(k)$, at which $x(t_1,k) = 1$. After this time
our general solution (\ref{genu}) takes the form,
\begin{eqnarray}
\lefteqn{u_{\rm ir}(t,k) \! \longrightarrow \! \frac{-i a^{\frac{D-1}2}}{
\sqrt{2k}} \frac{\Gamma(1 \!-\!\nu) J_{\!-\nu}(\!-\! \frac{x}{q})}{(\!-\!
\frac{x}{2 q})^{-\nu}} } \nonumber \\
& & \hspace{3cm} \times \Bigl(\frac{H_1}{k}\Bigr)^{\!\frac{D}2 \!-\! 1} 
{\cal C}_1(k) {\cal C}_i(k) .
\end{eqnarray}
Corrections are of order $x^2(t',k)$ times this result, where $t'$ is some
time for which $x(t',k) \ll 1$.

The factor ${\cal C}_1(k)$ depends upon the state of the system at first 
horizon crossing,
\begin{equation}
{\cal C}_1(k) \equiv \frac{\frac1{\sqrt{\pi}} \Gamma(\nu_1)}{(\!-\!
\frac1{2q})^{\nu_1 \!-\! \frac12}} \; .
\end{equation}
The factor ${\cal C}_i(k)$ depends upon evolution up to this time,
\begin{equation}
{\cal C}_i(k) \equiv {\cal M}^{11}(t_1,t_i,k) + \frac{i e^{-i \nu_i \pi}}{
\sin(\nu_i \pi)} {\cal M}^{12}(t_1,t_i,k) .
\end{equation}
Since both $y(t)$ and $\dot{y}(t)$ are small during inflation, we can obtain an
excellent approximation for ${\cal C}_i(k)$ by simply expanding the transfer 
matrix,
\begin{eqnarray}
\lefteqn{{\cal M}^{11}(t_1,t_i,k) = 1 + \int_{t_i}^{t_1} \!\!\! dt 
\gamma(t,k) } \nonumber \\
& & \!\!\!\!\! + \! \int_{t_i}^{t_1} \!\!\! dt \!\! \int_{t_i}^t \!\!\! dt' 
\Bigl[\gamma(t,k) \gamma(t',k) - \delta(t,k) \beta(t',k) \Bigr] + \dots \qquad 
\label{11comp} \\
\lefteqn{{\cal M}^{12}(t_1,t_i,k) = 0 -i\int_{t_i}^{t_1} \!\!\! dt \delta(t,k)
} \nonumber \\
& & \!\!\!\!\! -i \!\int_{t_i}^{t_1} \!\!\! dt \!\! \int_{t_i}^t \!\!\! dt' 
\Bigl[ \gamma(t,k) \delta(t',k) - \delta(t,k) \gamma(t',k) \Bigr] + \dots 
\label{12comp} \qquad
\end{eqnarray}
For completeness we remind the reader of the relations,
\begin{eqnarray}
\beta(t,k) & = & \frac{\pi \dot{\nu}}{2 \sin^2(\nu \pi)} b_{\nu}({\scriptstyle
\!-\! \frac{x}{q}}) ,\\
\gamma(t,k) & = & \frac{\pi \dot{\nu}}{4 \sin(\nu \pi)} c_{\nu}({\scriptstyle
\!-\! \frac{x}{q}}) ,\\
\delta(t,k) & = & \frac{\pi \dot{\nu}}{2} d_{\nu}({\scriptstyle \!-\! 
\frac{x}{q}}) .
\end{eqnarray}

The formalism we have just summarized has many applications. One of these is
to derive an improved estimate for the graviton contribution to the power
spectrum of anisotropies in the cosmic microwave background. We shall present
the derivation elsewhere \cite{TW2} but the result is,
\begin{equation}
{\cal P}_h(k) = G H_1^2(k) \Vert {\cal C}_i(k) \Vert^2 {\cal C}_1^2(k) \; .
\end{equation}
The standard result is $G H_1^2$, so the factors ${\cal C}_i(k)$ and ${\cal
C}_1(k)$ represent improvements. Because the literature abounds with different
conventions for this quantity we correspond ${\cal P}_h(k)$ below to the symbol
$\delta_h(k)$ used by Mukhanov, Feldman and Brandenberger \cite{MFB}, to
the variable ${\cal P}_g(k)$ used by Liddle and Lyth \cite{LL}, and to the
quantity $A_T^2(k)$ used by Lidsey {\it et al.} \cite{Rocky},
\begin{equation}
{\cal P}_h(k) = \frac{9 \pi}4 \Vert \delta_h(k) \Vert^2 = \frac{\pi}{16} 
{\cal P}_g(k) = \frac{25 \pi}4 A_T^2(k) \; .
\end{equation}
The most unambiguous definition of ${\cal P}_h(k)$ is given by the (purely
gravitational wave contribution to the) correlation function between 
temperature fluctuations observed from directions $\widehat{e}_1$ and 
$\widehat{e}_2$,
\begin{eqnarray}
\!\!\lefteqn{\Bigl\langle \Omega \Bigl\vert \frac{\Delta T_R(\widehat{e}_1)}{
T_R} \frac{\Delta T_R(\widehat{e}_2)}{T_R} \Bigr\vert \Omega \Bigr\rangle \!= 
\!\! \int_0^{\infty} \!\! \frac{dk}k {\cal P}_h(k) \int \!\! \frac{d^2
\widehat{k}}{4 \pi} \Theta(\widehat{e}_1,\vec{k}) } \nonumber \\
& & \hspace{.5cm} \times \Theta^*(\widehat{e}_2,\vec{k}) \widehat{e}^{\, i}_1 
\widehat{e}^{\,j}_1 \widehat{e}^{\, m}_2 \widehat{e}^{\, n}_2 \Bigl[\Pi_{im} 
\Pi_{jn} \!-\! \frac12 \Pi_{ij} \Pi_{mn} \Bigr] . \qquad
\end{eqnarray}
In this formula $\Pi_{ij} \equiv \delta_{ij} - \widehat{k}_i \widehat{k}_j$
is the transverse projection matrix coming from the sum over graviton
polarization tensors. The angular factor is,
\begin{equation}
\Theta(\widehat{e},\vec{k}) = \frac{e^{2 i \theta \, x_0}}{\sqrt{4 \pi}} 
\Bigl\{2 \!-\! 3 \theta^2 \!-\! \frac32 \theta (1 \! - \! \theta^2) 
\Bigl[\ln\Bigl(\frac{1 \!+\! \theta}{1 \!-\! \theta}\Bigr) \!+\! i \pi \Bigr] 
\!\Bigr\} , \qquad
\end{equation}
where $x_0 \equiv x(t_0,k)$ is the physical wave number in current Hubble
units and we define $\theta \equiv \widehat{k} \cdot \widehat{e}$.

There are many other applications. One particularly interesting question we now
have the technology to answer is the response to multiple periods of inflation.
Whereas the weak energy condition implies that the Hubble parameter can only
decrease or stay constant, the evolution of $q(t)$ need not consist of
monotonic growth from its inflationary value of nearly $-1$. Indeed, after
inflation we know that $q(t)$ decreased from the era of radiation domination
($q=+1$) to the era of matter domination ($q=+\frac12$), and considerable
evidence exists that it has now dropped below zero \cite{SNa,SNb,WMAP}. There
is no reason it could not have experienced such oscillations at very early
times. The standard formalism for analyzing density perturbations breaks
down in this case, but our techniques should be applicable to any number of
horizon crossings.

Note that our formalism depends only upon the scale factor $a(t)$, not on 
the matter theory which supports it. Results derived in this way are 
therefore independent of the details of particular inflationary models. 
For example, it does not matter how many scalars, if any, participate in 
driving inflation.

It should be possible to compute the one loop effective action of the 
massless, minimally coupled scalar as a functional of general $a(t)$. Of 
course the ultraviolet divergence is universal and was derived long ago 
\cite{BD}, but the ultraviolet finite, infrared contributions have never 
been computed and are, in many ways more interesting. Our infrared expression
(\ref{IRu}) for the mode function should be of great use there.

Another application is computing the ultraviolet finite part of the 
coincidence limit of the propagator. Since linearized gravitons 
have the same mode functions, this would also apply for quantum gravity.
Proceeding in this way to compute the infrared part of the non-coincident
propagator, it should be possible to derive the quantum gravitationally 
induced stress tensor in the presence of a class of backgrounds guaranteed 
to include any solution which preserves homogeneity and isotropy. This is
an essential step in checking whether quantum gravitational back-reaction
can quench $\Lambda$-driven inflation \cite{TW1}.

Finally, a simple modification of the formalism developed here can be used
to derive the mode functions for general perturbations in gravity plus a
single, minimally coupled scalar \cite{TW3}. These mode functions can be
used to improve the scalar power spectrum as was done for its tensor
counterpart \cite{TW2}.

{\it 9. Acknowledgements:} This work was partially supported by EU
grants HPRN-CT-2000-00122 and HPRN-CT-2000-00131, by DOE contract 
DE-FG02-97ER41029 and by the Institute for Fundamental Theory.


\end{document}